\documentclass[aps,prl,preprint,tightenlines,superscriptaddress,showpacs,byrevtex,floatfix]{revtex4}
\usepackage{graphicx,color} 
\usepackage{dcolumn}  
\usepackage{amsmath}  
\usepackage{refmerge}

\begin{document}


\def\LR{\mathcal{R}}
\def\dE{{\Delta E}}
\def\mb{{M_{\rm bc}}}
\def\mkpi{{M_{K\pi}}}
\def\mpipi{{M_{\pi\pi}}}
\def\Dt{\Delta t}
\def\Dz{\Delta z}
\def\egcms{{E_\gamma^{\rm c.m.s.}}}
\newcommand{\dmd}{\Delta m_d}
\def\fol{f_{\rm ol}}
\def\fr{f_{\rhoz\gamma}}
\def\fk{f_{\kstarz\gamma}}
\def\fbb{f_{\BB}}

\newcommand{\fCP}{f_{\rm rec}}
\newcommand{\zCP}{z_{\rm rec}}
\newcommand{\tCP}{t_{\rm rec}}
\newcommand{\ftag}{f_{\rm tag}}
\newcommand{\ttag}{t_{\rm tag}}
\newcommand{\ztag}{z_{\rm tag}}
\newcommand{\cala}{{\mathcal A}}
\newcommand{\cals}{{\mathcal S}}
\newcommand{\calb}{{\mathcal B}}

\def\rsigbkg{{\mathcal R}_{\rm s/b}}
\def\rsigbkgBH{{\mathcal R}_{\rm s/b}^{\rm BH}}
\def\Rrho{R_{\rhoz\gamma}}
\def\Rbkg{R_{\qq}}
\def\calf{{\mathcal F}}
\def\taubz{{\tau_\bz}}
\def\taubp{{\tau_\bp}}
\def\tauc{{\tau_{\ks\pip\gamma}}}
\def\taun{{\tau_{\ks\piz\gamma}}}
\newcommand*{\fq}{\ensuremath{q}}

\def\bz{{B^0}}
\def\bzb{{\overline{B}{}^0}}
\def\bbar{{\overline{B}}}
\def\bp{{B^+}}
\def\bm{{B^-}}
\def\kz{{K^0}}
\def\ks{{K_S^0}}
\def\kp{{K^+}}
\def\km{{K^-}}
\def\rhoz{{\rho^0}}
\def\pip{{\pi^+}}
\def\pim{{\pi^-}}
\def\piz{{\pi^0}}
\def\kstar{{K^\ast}}
\def\kstarz{{K^{\ast 0}}}
\def\kstarp{{K^{\ast +}}}
\def\kstarm{{K^{\ast -}}}
\def\ktwostar{{K_2^\ast}}
\def\ktwostarz{{K_2^{\ast 0}}}
\def\kstarpm{{K^{\ast\pm}}}
\def\jpsi{{J/\psi}}
\def\qq{{q\bar{q}}}
\def\BB{{B\bar{B}}}

\def\GeV{\,{\rm GeV}}
\def\GeVc{\,{\rm GeV}/c}
\def\GeVcc{\,{\rm GeV}/c^2}
\def\MeV{\,{\rm MeV}}
\def\MeVcc{\,{\rm MeV}/c^2}

\newcommand{\BaBar}{{\sc B\hspace*{-0.2ex}a\hspace*{-0.2ex}B\hspace*{-0.2ex}a\hspace*{-0.2ex}r}}

\def\NBBsvdI{152\times 10^6}
\def\NBBothree{275\times 10^6}
\def\NBBofour{388\times 10^6}
\def\NBBofive{535\times 10^6}
\def\NBBosix{575\times 10^6}
\def\NBBoseven{657\times 10^6}

\def\NBBsvdII{505\times 10^6}

\newcommand{\asyme}[3]{{#1^{+#2}_{-#3}}}
\newcommand{\syme}[2]{{{#1}\pm {#2}}}

\newcommand{\symeSS}[2]{{{#1}\pm {#2}\mbox{(stat)}}}

\newcommand{\asymasyme}[5]{{#1^{+#2}_{-#3}{}^{+#4}_{-#5}}}
\newcommand{\symasyme}[4]{{{#1}\pm {#2}{}^{+#3}_{-#4}}}
\newcommand{\asymsyme}[4]{{{#1}^{+#2}_{-#3}\pm {#4}}}
\newcommand{\symsyme}[3]{{{#1}\pm {#2}\pm {#3}}}

\newcommand{\asymasymeSS}[5]{{#1^{+#2}_{-#3}\mbox{(stat)}{}^{+#4}_{-#5}\mbox{(syst)}}}
\newcommand{\symasymeSS}[4]{{{#1}\pm {#2}\mbox{(stat)}{}^{+#3}_{-#4}\mbox{(syst)}}}
\newcommand{\asymsymeSS}[4]{{{#1}^{+#2}_{-#3}\mbox{(stat)}\pm {#4}\mbox{(syst)}}}
\newcommand{\symsymeSS}[3]{{{#1}\pm {#2}\mbox{(stat)}\pm {#3}\mbox{(syst)}}}


\def\SkspizgmValBelle{-0.10}
\def\SkspizgmStatBelle{0.31}
\def\SkspizgmSystBelle{0.07}
\def\AkspizgmValBelle{-0.20}
\def\AkspizgmStatBelle{0.20}
\def\AkspizgmSystBelle{0.06}
\def\SkspizgmBelle{\symsyme{\SkspizgmValBelle}{\SkspizgmStatBelle}{\SkspizgmSystBelle}}
\def\SkspizgmBelleSS{\symsymeSS{\SkspizgmValBelle}{\SkspizgmStatBelle}{\SkspizgmSystBelle}}
\def\AkspizgmBelle{\symsyme{\AkspizgmValBelle}{\AkspizgmStatBelle}{\AkspizgmSystBelle}}
\def\AkspizgmBelleSS{\symsymeSS{\AkspizgmValBelle}{\AkspizgmStatBelle}{\AkspizgmSystBelle}}
\def\SkstgmValBelle{-0.32}
\def\SkstgmStatpBelle{0.36}
\def\SkstgmStatmBelle{0.33}
\def\SkstgmSystBelle{0.05}
\def\AkstgmValBelle{-0.20}
\def\AkstgmStatBelle{0.24}
\def\AkstgmSystBelle{0.05}
\def\SkstgmBelle{\asymsyme{\SkstgmValBelle}{\SkstgmStatpBelle}{\SkstgmStatmBelle}{\SkstgmSystBelle}}
\def\SkstgmBelleSS{\asymsymeSS{\SkstgmValBelle}{\SkstgmStatBelle}{\SkstgmSystBelle}}
\def\AkstgmBelle{\symsyme{\AkstgmValBelle}{\AkstgmStatBelle}{\AkstgmSystBelle}}
\def\AkstgmBelleSS{\symsymeSS{\AkstgmValBelle}{\AkstgmStatBelle}{\AkstgmSystBelle}}

\def\SkspizgmValBaBar{-0.06}
\def\SkspizgmStatBaBar{0.37}
\def\AkspizgmValBaBar{0.48}
\def\AkspizgmStatBaBar{0.22}
\def\SkspizgmBaBar{\syme{\SkspizgmValBaBar}{\SkspizgmStatBaBar}}
\def\SkspizgmBaBarSS{\symeSS{\SkspizgmValBaBar}{\SkspizgmStatBaBar}}
\def\AkspizgmBaBar{\syme{\AkspizgmValBaBar}{\AkspizgmStatBaBar}}
\def\AkspizgmBaBarSS{\symeSS{\AkspizgmValBaBar}{\AkspizgmStatBaBar}}
\def\SkstgmValBaBar{-0.21}
\def\SkstgmStatBaBar{0.40}
\def\SkstgmSystBaBar{0.05}
\def\AkstgmValBaBar{0.40}
\def\AkstgmStatBaBar{0.23}
\def\AkstgmSystBaBar{0.04}
\def\SkstgmBaBar{\asymsyme{\SkstgmValBaBar}{\SkstgmStatpBaBar}{\SkstgmStatmBaBar}{\SkstgmSystBaBar}}
\def\SkstgmBaBarSS{\asymsymeSS{\SkstgmValBaBar}{\SkstgmStatBaBar}{\SkstgmSystBaBar}}
\def\AkstgmBaBar{\symsyme{\AkstgmValBaBar}{\AkstgmStatBaBar}{\AkstgmSystBaBar}}
\def\AkstgmBaBarSS{\symsymeSS{\AkstgmValBaBar}{\AkstgmStatBaBar}{\AkstgmSystBaBar}}

\def\BFrhozgmValBelle{1.25}
\def\BFrhozgmStatpBelle{0.37}
\def\BFrhozgmStatmBelle{0.33}
\def\BFrhozgmSystpBelle{0.07}
\def\BFrhozgmSystmBelle{0.06}
\def\BFrhozgmBelle{\asymasyme{\BFrhozgmValBelle}{\BFrhozgmStatpBelle}{\BFrhozgmStatmBelle}{\BFrhozgmSystpBelle}{\BFrhozgmSystmBelle}}
\def\BFrhozgmBelleSS{\asymasymeSS{\BFrhozgmValBelle}{\BFrhozgmStatpBelle}{\BFrhozgmStatmBelle}{\BFrhozgmSystpBelle}{\BFrhozgmSystmBelle}}

\def\BFomegagmValBelle{0.56}
\def\BFomegagmStatpBelle{0.34}
\def\BFomegagmStatmBelle{0.27}
\def\BFomegagmSystpBelle{0.05}
\def\BFomegagmSystmBelle{0.10}
\def\BFomegagmBelle{\asymasyme{\BFomegagmValBelle}{\BFomegagmStatpBelle}{\BFomegagmStatmBelle}{\BFomegagmSystpBelle}{\BFomegagmSystmBelle}}
\def\BFomegagmBelleSS{\asymasymeSS{\BFomegagmValBelle}{\BFomegagmStatpBelle}{\BFomegagmStatmBelle}{\BFomegagmSystpBelle}{\BFomegagmSystmBelle}}

\def\BFrhozgmValBaBar{0.79}
\def\BFrhozgmStatpBaBar{0.22}
\def\BFrhozgmStatmBaBar{0.20}
\def\BFrhozgmSystBaBar{0.06}
\def\BFrhozgmBaBar{\asymsyme{\BFrhozgmValBaBar}{\BFrhozgmStatpBaBar}{\BFrhozgmStatmBaBar}{\BFrhozgmSystBaBar}}
\def\BFrhozgmBaBarSS{\asymsymeSS{\BFrhozgmValBaBar}{\BFrhozgmStatpBaBar}{\BFrhozgmStatmBaBar}{\BFrhozgmSystBaBar}}

\def\BFomegagmValBaBar{0.40}
\def\BFomegagmStatpBaBar{0.24}
\def\BFomegagmStatmBaBar{0.20}
\def\BFomegagmSystBaBar{0.05}
\def\BFomegagmBaBar{\asymsyme{\BFomegagmValBaBar}{\BFomegagmStatpBaBar}{\BFomegagmStatmBaBar}{\BFomegagmSystBaBar}}
\def\BFomegagmBaBarSS{\asymsymeSS{\BFomegagmValBaBar}{\BFomegagmStatpBaBar}{\BFomegagmStatmBaBar}{\BFomegagmSystBaBar}}


\def\NrhozgammaCandInFit{5362}
\def\NrhozgammaCandInBox{410}
\def\Nrhozgam{\syme{48.3}{13.5}}
\def\Nkstarzgam{\syme{180.6}{16.8}}
\def\NBB{\syme{10.3}{4.3}}
\def\Nqq{\syme{168.8}{2.6}}

\def\pdglife{\syme{1.530}{0.009}}

\def\efflifer{\syme{1.26}{0.06}}
\def\efflifek{\syme{1.40}{0.02}}

\def\rlife{\asyme{1.26}{0.37}{0.30}}
\def\klife{\syme{1.54}{0.16}}
\def\SrhozgmVal{-0.83}
\def\SrhozgmStat{0.65}
\def\SrhozgmSyst{0.18}
\def\ArhozgmVal{-0.44}
\def\ArhozgmStat{0.49}
\def\ArhozgmSyst{0.14}
\def\SrhozgmResult{\symsyme{\SrhozgmVal}{\SrhozgmStat}{\SrhozgmSyst}}
\def\SrhozgmResultSS{\symsymeSS{\SrhozgmVal}{\SrhozgmStat}{\SrhozgmSyst}}
\def\ArhozgmResult{\symsyme{\ArhozgmVal}{\ArhozgmStat}{\ArhozgmSyst}}
\def\ArhozgmResultSS{\symsymeSS{\ArhozgmVal}{\ArhozgmStat}{\ArhozgmSyst}}

\def\SkrVal{+0.02}
\def\SkrStat{0.25}
\def\AkrVal{-0.03}
\def\AkrStat{0.17}
\def\SkrResult{\syme{\SkrVal}{\SkrStat}}
\def\AkrResult{\syme{\AkrVal}{\AkrStat}}

\def\controllife{\syme{1.57}{0.04}}

\def\SkstzgmVal{+0.05}
\def\SkstzgmStat{0.07}
\def\AkstzgmVal{-0.01}
\def\AkstzgmStat{0.05}
\def\SkstzgmResult{\syme{\SkstzgmVal}{\SkstzgmStat}}
\def\AkstzgmResult{\syme{\AkstzgmVal}{\AkstzgmStat}}

\preprint{\vbox{ \hbox{   }
    \hbox{Belle Preprint 2007-40}
    \hbox{KEK Preprint 2007-43}
  }
}

\title{ \quad\\[0.5cm] Time-Dependent {\boldmath $CP$}-Violating
Asymmetry in $\bz\to\rhoz\gamma$ Decays\\
}
\date{\today}

\affiliation{Budker Institute of Nuclear Physics, Novosibirsk}
\affiliation{Chiba University, Chiba}
\affiliation{University of Cincinnati, Cincinnati, Ohio 45221}
\affiliation{The Graduate University for Advanced Studies, Hayama}
\affiliation{Hanyang University, Seoul}
\affiliation{University of Hawaii, Honolulu, Hawaii 96822}
\affiliation{High Energy Accelerator Research Organization (KEK), Tsukuba}
\affiliation{Hiroshima Institute of Technology, Hiroshima}
\affiliation{Institute of High Energy Physics, Chinese Academy of Sciences, Beijing}
\affiliation{Institute of High Energy Physics, Vienna}
\affiliation{Institute of High Energy Physics, Protvino}
\affiliation{Institute for Theoretical and Experimental Physics, Moscow}
\affiliation{J. Stefan Institute, Ljubljana}
\affiliation{Kanagawa University, Yokohama}
\affiliation{Korea University, Seoul}
\affiliation{Kyoto University, Kyoto}
\affiliation{Kyungpook National University, Taegu}
\affiliation{\'Ecole Polytechnique F\'ed\'erale de Lausanne (EPFL), Lausanne}
\affiliation{University of Ljubljana, Ljubljana}
\affiliation{University of Maribor, Maribor}
\affiliation{University of Melbourne, School of Physics, Victoria 3010}
\affiliation{Nagoya University, Nagoya}
\affiliation{Nara Women's University, Nara}
\affiliation{National Central University, Chung-li}
\affiliation{National United University, Miao Li}
\affiliation{Department of Physics, National Taiwan University, Taipei}
\affiliation{H. Niewodniczanski Institute of Nuclear Physics, Krakow}
\affiliation{Nippon Dental University, Niigata}
\affiliation{Niigata University, Niigata}
\affiliation{University of Nova Gorica, Nova Gorica}
\affiliation{Osaka City University, Osaka}
\affiliation{Osaka University, Osaka}
\affiliation{Panjab University, Chandigarh}
\affiliation{Saga University, Saga}
\affiliation{University of Science and Technology of China, Hefei}
\affiliation{Seoul National University, Seoul}
\affiliation{Sungkyunkwan University, Suwon}
\affiliation{University of Sydney, Sydney, New South Wales}
\affiliation{Toho University, Funabashi}
\affiliation{Tohoku Gakuin University, Tagajo}
\affiliation{Tohoku University, Sendai}
\affiliation{Department of Physics, University of Tokyo, Tokyo}
\affiliation{Tokyo Institute of Technology, Tokyo}
\affiliation{Tokyo Metropolitan University, Tokyo}
\affiliation{Tokyo University of Agriculture and Technology, Tokyo}
\affiliation{Virginia Polytechnic Institute and State University, Blacksburg, Virginia 24061}
\affiliation{Yonsei University, Seoul}
  \author{Y.~Ushiroda}\affiliation{High Energy Accelerator Research Organization (KEK), Tsukuba} 
  \author{K.~Sumisawa}\affiliation{High Energy Accelerator Research Organization (KEK), Tsukuba} 
  \author{N.~Taniguchi}\affiliation{Kyoto University, Kyoto} 
  \author{I.~Adachi}\affiliation{High Energy Accelerator Research Organization (KEK), Tsukuba} 
  \author{H.~Aihara}\affiliation{Department of Physics, University of Tokyo, Tokyo} 
  \author{K.~Arinstein}\affiliation{Budker Institute of Nuclear Physics, Novosibirsk} 
  \author{T.~Aushev}\affiliation{\'Ecole Polytechnique F\'ed\'erale de Lausanne (EPFL), Lausanne}\affiliation{Institute for Theoretical and Experimental Physics, Moscow} 
  \author{S.~Bahinipati}\affiliation{University of Cincinnati, Cincinnati, Ohio 45221} 
  \author{A.~M.~Bakich}\affiliation{University of Sydney, Sydney, New South Wales} 
  \author{V.~Balagura}\affiliation{Institute for Theoretical and Experimental Physics, Moscow} 
% ...
\collaboration{The Belle Collaboration}


\begin{abstract}
 We report the first measurement of $CP$-violation parameters in
 $\bz\to\rhoz\gamma$ decays based on a data sample of $\NBBoseven\,\BB$ pairs
 collected with the Belle detector at the KEKB asymmetric-energy
 $e^+\,e^-$ collider. We obtain the
 time-dependent and direct $CP$-violating parameters,
 $\cals_{\rhoz\gamma} = \SrhozgmResultSS$ and
 $\cala_{\rhoz\gamma} = \ArhozgmResultSS$, respectively.
\end{abstract}


\maketitle

\begin{figure}
 \resizebox{0.6\columnwidth}{!}{\includegraphics{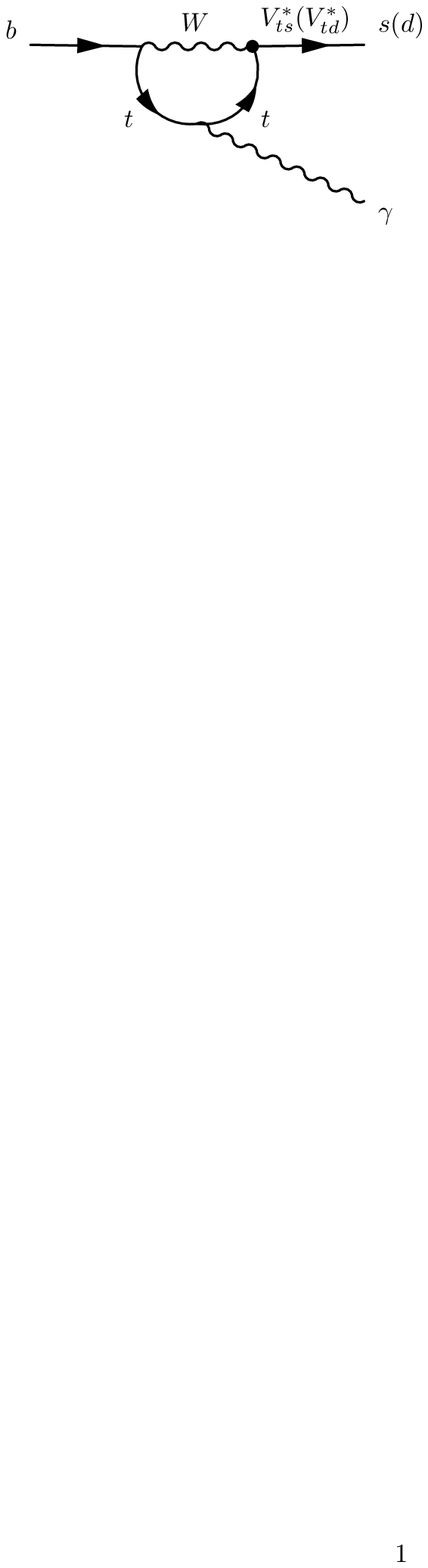}}
 \caption{Feynman diagram for radiative $b$ decay in the SM.}
 \label{fig:feyn}
\end{figure}

Radiative decay processes are sensitive to physics beyond the standard
model (SM). Figure~\ref{fig:feyn} shows the lowest order Feynman
diagram for radiative $b$ decay in the SM. The heavy SM particles in the
loop can be replaced by new physics (NP) particles.
Hence the corresponding physics observables may deviate from SM expectations.
Recently, the possibility of time-dependent $CP$ asymmetries in $b\to s\gamma$
from NP have drawn much theoretical and experimental
interest~\cite{Atwood:1997zr,Ball:2006eu,Ushiroda:2006fi,Aubert:2005bu}. Both
Belle~\cite{Ushiroda:2006fi} and \BaBar~\cite{Aubert:2005bu} have
measured time-dependent $CP$-violating parameters in
$\bz\to\ks\piz\gamma$ decay.  The results so far are consistent with the SM.

Signals for $\bz\to\rhoz\gamma$ have been established by
Belle~\cite{Mohapatra:2005rj} and \BaBar~\cite{Aubert:2006pu},
which enables us to measure $CP$ asymmetries in the $b\to d\gamma$
process.
As in $b\to s\gamma$, the photon emitted in $b\to d\gamma$
($\bar{b}\to\bar{d}\gamma$) is predominantly left-handed (right-handed),
and hence the final state is flavor specific~\cite{Atwood:1997zr}.
In the decay $\bz\to\rhoz\gamma$, the SM predicts no time-dependent
$CP$ asymmetry ($\cals$) and $-0.1$ for the direct $CP$
asymmetry ($\cala$)~\cite{Ball:2006eu,Ali:2004hn}.
In particular, assuming the top quark is the dominant contribution in
the loop shown in Fig.~\ref{fig:feyn},
the decay amplitude has a weak phase $\phi_1$ that cancels the phase in
the mixing; consequently $\cals$ vanishes.
Observing a non-zero value of $\cals$ would indicate effects of
NP~\cite{Kim:2004zm}.
In this Letter, we present the first measurements of $\cals$ and $\cala$
for the $\bz\to\rhoz(\to\pip\pim)\gamma$ transition based on $\NBBoseven\,\BB$
pairs collected with the Belle detector~\cite{Belle} at the KEKB
asymmetric-energy $e^+\,e^-$ (3.5 on 8.0$\GeV$) collider~\cite{bib:KEKB}.

The Belle detector is a large-solid-angle magnetic
spectrometer that
consists of a silicon vertex detector (SVD),
a 50-layer central drift chamber, an array of
aerogel threshold Cherenkov counters, 
a barrel-like arrangement of time-of-flight
scintillation counters, and an electromagnetic calorimeter (ECL)
comprised of CsI(Tl) crystals located inside 
a superconducting solenoid coil that provides a 1.5\,T
magnetic field.  An iron flux-return located outside of
the coil is instrumented to detect $K_L^0$ mesons and to identify
muons.

At the KEKB, the
$\Upsilon(4S)$ is produced with a Lorentz
boost of $\beta\gamma=0.425$ along the $z$ axis, which is defined as the
direction antiparallel to the $e^+$ beam direction.
In the decay chain $\Upsilon(4S)\to \bz\bzb \to \fCP \ftag$, where one
of the $B$ mesons decays at time $\tCP$ to a final state $\fCP$, which
is our signal mode, and the other decays at time $\ttag$ to a final
state $\ftag$ that distinguishes between $B^0$ and $\bzb$, the decay
rate has a time dependence given by
\begin{equation}
 \begin{split}
  {\cal P}(\Dt)& =
  \frac{e^{-|\Dt|/{\taubz}}}{4{\taubz}}
  \biggl\{1+\fq
  \Bigl[ \cals\sin(\dmd\Dt)\\
  &
  + \cala\cos(\dmd\Dt)
  \Bigr]
  \biggr\}.
  \label{eq:psig}
 \end{split}
\end{equation}
Here $\taubz$ is the
$B^0$ lifetime, $\dmd$ is the mass difference between the two $B^0$ mass
eigenstates, $\Dt$ is the time difference $\tCP - \ttag$, and the
$b$-flavor charge $\fq$ = +1 ($-1$) when the tagging $B$ meson is a
$B^0$ ($\bzb$).
Since the $B^0$ and $\bzb$ mesons are approximately at 
rest in the $\Upsilon(4S)$ center-of-mass system (c.m.s.),
$\Dt$ can be determined from the displacement in $z$ 
between the $\fCP$ and $\ftag$ decay vertices:
$\Delta t \simeq (\zCP - \ztag)/(\beta\gamma c)
 \equiv \Delta z/(\beta\gamma c)$.

We reconstruct
$\bz\to\rhoz\gamma$, as well as a control sample of
$\bz\to\kstarz(\to\kp\pim)\gamma$~\cite{bib:CC}.
For high energy prompt photons, we select the cluster in the ECL with
the highest energy in the c.m.s. from clusters that have no associated
charged track.  We require $1.4\GeV < \egcms < 3.4\GeV$.  For the
selected photon, we also require $E_9/E_{25}>0.95$, where $E_9/E_{25}$
is the ratio of energies summed in $3\times 3$ and $5\times 5$ arrays of
CsI(Tl) crystals around the center of the shower. In order to reduce the
background from $\piz\to\gamma\gamma$ or $\eta\to\gamma\gamma$ decays,
photons from these decays are rejected as described
in~\cite{Koppenburg:2004fz}; this retains 97\% of the signal and rejects
 20\% of the background events. The polar angle of the photon direction in
the laboratory frame is restricted to the barrel region of the ECL
($33^\circ < \theta_\gamma < 128^\circ$).

Charged tracks are required to originate from the vicinity of the interaction point (IP),
within 3\,cm in $z$ and 0.5\,cm in $r$-$\phi$; their transverse
momenta are required to be greater than $0.22\GeVc$.
Charged tracks from $\ks$ decays as well as positively identified
protons, muons and electrons are excluded.
Finally, candidate tracks are classified as pion candidates and kaon candidates
according to the ratio of kaon and pion particle identification likelihoods.
This selection retains 87\% of pions while rejecting 92\% of kaons.
Pairs of oppositely charged pions are combined to form $\rhoz$ candidates.
Oppositely charged kaon and pion candidates are combined to form $\kstarz$
candidates.
We form the invariant mass $\mkpi$ for $\kstarz$ and $\rhoz$ candidates.
To obtain $\mkpi$ for $\rhoz$ candidates,
we assign the kaon mass to
each pion in turn,
and take the lower of the two values.
We use $\mkpi$ rather than $\mpipi$ since it gives a better separation
of the $\rhoz\gamma$ signal from the $\kstarz\gamma$ background.

We form two kinematic variables: the energy difference
$\dE\equiv (\sum_i E_i^\ast)-E_{\rm beam}^\ast$
and the beam-energy constrained mass
$\mb\equiv\sqrt{(E_{\rm beam}^\ast)^2-(\sum_ip_i^\ast)^2}$,
where $E_{\rm beam}^\ast$ is the beam energy in the c.m.s.,
$E_i^\ast$ and $p_i^\ast$ are the energy and momentum of the $i$-th final state
particle in the c.m.s., and the summation is taken over all the final state
particles of the candidate $B$ meson.
Unlike $\mkpi$, we do not assign the kaon mass but instead assign the pion mass
to form the energy and the momentum of $\rhoz\gamma$ candidates.
The signal box in $\dE$, $\mb$ and $\mkpi$, which is used for the
measurements of $CP$-violating parameters,
is defined as $-0.15\GeV \le \dE \le 0.1\GeV$,
$5.27\GeVcc \le \mb \le 5.29\GeVcc$ and $0.7\GeVcc < \mkpi < 1.1\GeVcc$.
A larger region in $\dE$ and $\mb$, $-0.3\GeV < \dE < 0.5\GeV$ and
$5.2\GeVcc < \mb$ is used to determine the signal and background fractions.

In order to suppress the background contribution from $\qq$
($e^+e^-\to\qq$ with $q = u,d,s,c$),
an event likelihood ratio $\LR$ is formed from likelihood variables for
signal ($\mathcal{L}_{\rm sig}$) and background ($\mathcal{L}_{\rm bkg}$) as
$\LR\equiv\mathcal{L}_{\rm sig}/(\mathcal{L}_{\rm sig}+\mathcal{L}_{\rm
bkg})$.
These likelihood variables are obtained by combining three variables:
a Fisher discriminant $\calf$~\cite{Fisher} that uses modified Fox-Wolfram
moments~\cite{Abe:2003yy} as discriminating variables, the polar angle
of the $B$ meson candidate momentum in the c.m.s. ($\cos\theta_B$),
and the cosine of the helicity angle ($\cos\theta_H$)
defined as the momentum direction of the $\pip$ with respect to the opposite
of the $B$ momentum in the $\rhoz$ rest frame
(similary for $\kstarz\gamma$).
We also require $|\cos\theta_H|<0.75$ in order to suppress background
from random low momentum pions.
$\LR$ is also used to determine the best candidate when multiple
candidates are found in a single event, although the fraction of events
with multiple candidates is small (0.7\%).

There is a large background from $\bz\to\kstarz\gamma$, which has
a branching fraction forty times larger than that of $\bz\to\rhoz\gamma$.
When a kaon is misidentified as a pion, the $\kstarz\gamma$ events easily
mimic the $\rhoz\gamma$ signal. This background peaks at $\kstarz$ mass
in $\mkpi$, and distributes in low $\dE$ region because the pion mass is
assigned to the kaon.
However, this is still acceptable since the $CP$ asymmetries
in the $\bz\to\kstarz\gamma$ decay are known with good precision.
There are several background contributions from $B$ decays that could
have finite $CP$ asymmetries,
$\rho^+\pim$, $\rhoz\piz$, and $\pip\pim\eta$; however the contributions
from these modes are small and thus their impact on our measurement is
tiny.

The $b$-flavor of the accompanying $B$ meson is identified from
inclusive properties of particles that are not associated with the
reconstructed signal decay.  The algorithm for flavor tagging is
described in detail elsewhere~\cite{bib:fbtg_nim}.  We use two
parameters, $\fq$ defined in Eq.~(\ref{eq:psig}) and $r$, to represent
the tagging information.  The parameter $r$ is an event-by-event
flavor-tagging quality factor that ranges from 0 to 1: $r=0$ when there
is no flavor discrimination and $r=1$ when the flavor assignment is
unambiguous. The value of $r$ is determined by using Monte Carlo (MC)
and is used to sort data
into seven $r$ intervals. Events with $r>0.1$ are sorted into six $r$
intervals; for each interval, the wrong-tag fraction $w$ and the 
difference $\Delta w$ in $w$ between the $\bz$ and $\bzb$ decays are
determined from high-statistics control samples of semi-leptonic
and hadronic $b\to c$ decays. For events with $r\le 0.1$,
there is negligible flavor discrimination available and we set $w$ to
0.5.

The vertex position of the signal-side decay of $\bz\to\rhoz\gamma$ and
the control sample $\bz\to\kstarz\gamma$ is reconstructed from one or two
charged track trajectories that have enough hits in the SVD,
with a constraint on the IP.
The IP profile
($\sigma_x\simeq 100\rm\,\mu m$, $\sigma_y\simeq 5\rm\,\mu m$) is
smeared by the finite $B$
flight length in the plane perpendicular to the $z$ axis ($21\,\mu m$).
The other (tag-side) $B$ vertex is determined from well reconstructed
tracks that are not assigned to the signal side.  A constraint to the IP
profile is also imposed. The resolution of the distance of the two $B$
vertices is typically $160\rm\,\mu m$.

After all the selections are applied, we obtain $\NrhozgammaCandInFit$
candidates in the $\dE$-$\mb$-$\mkpi$ fit region, of which
$\NrhozgammaCandInBox$ are in the signal box. We perform an extended
unbinned maximum likelihood (UML) fit to the $\dE$-$\mb$-$\mkpi$
distribution in order to resolve the $\rhoz\gamma$, $\kstarz\gamma$,
other $\BB$ and $\qq$ components.

The probability density function (PDF) for $\rhoz\gamma$ and
$\kstarz\gamma$ are obtained from MC.  We use a two-dimensional
histogram for $\mb$-$\dE$, and two one-dimensional histograms for
$\mkpi$ depending on $\dE$. For these PDFs, the peak position and the
width are corrected using the $\bz\to\kstarz\gamma$ control sample in
order to account for differences between data and simulation.
The PDF for the other $\BB$ background component,
which populates the lower $\dE$ region,
is also obtained from MC. For $\qq$ background, we use the
product of one dimensional PDFs: the ARGUS
parameterization~\cite{bib:ARGUS} for $\mb$, a first-order polynomial
for $\dE$, and a 20 bin histogram for $\mkpi$. The shape parameters (one
ARGUS coefficient, one polynomial coefficient, and fractions of 19 bin
contents) are determined in the fit. Together with the yield of the four
components, we have 25 free parameters in the fit.

\begin{figure}
\begin{tabular}{cc}
 \resizebox{0.49\columnwidth}{!}{\includegraphics{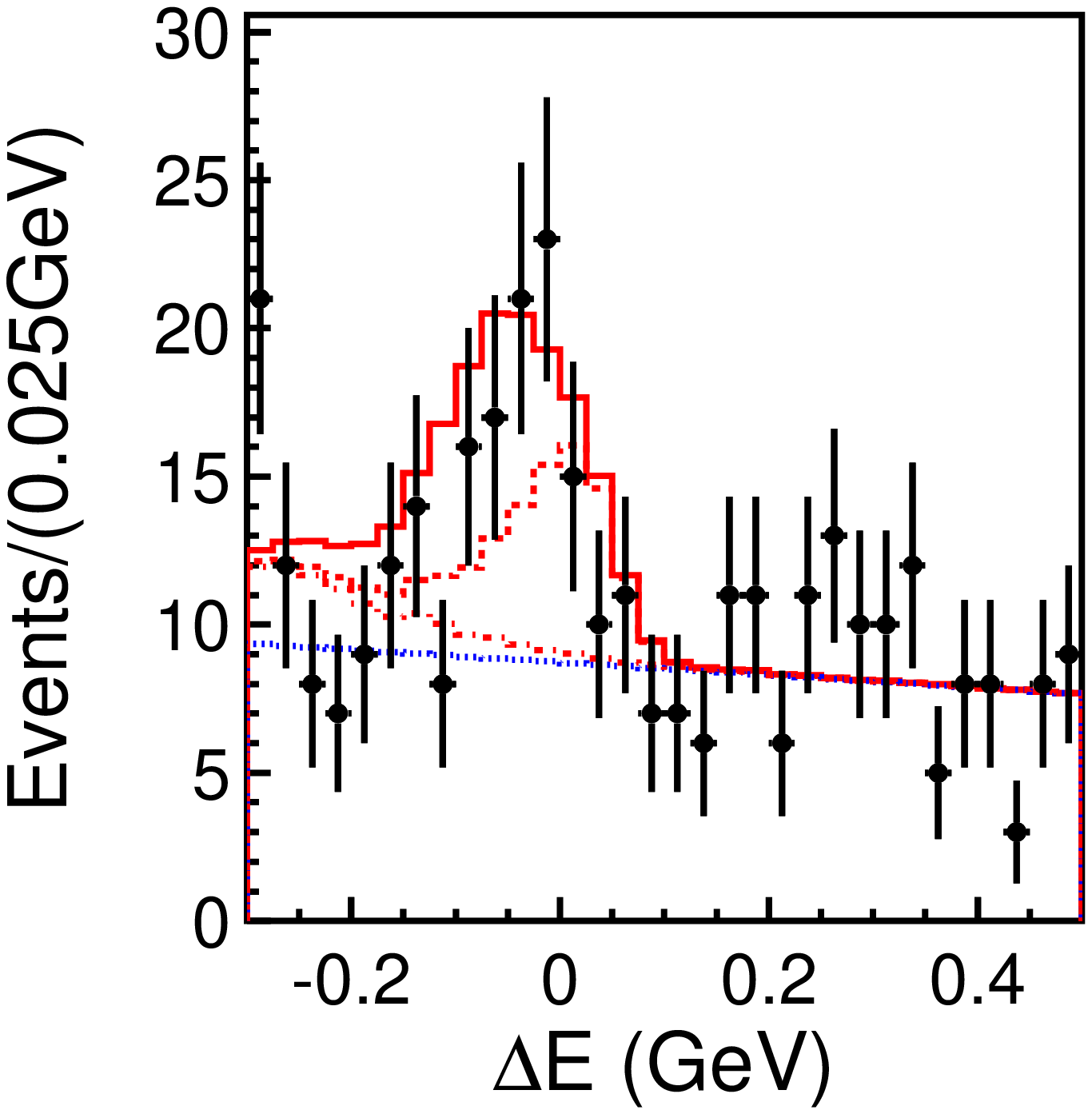}}
 &
 \resizebox{0.49\columnwidth}{!}{\includegraphics{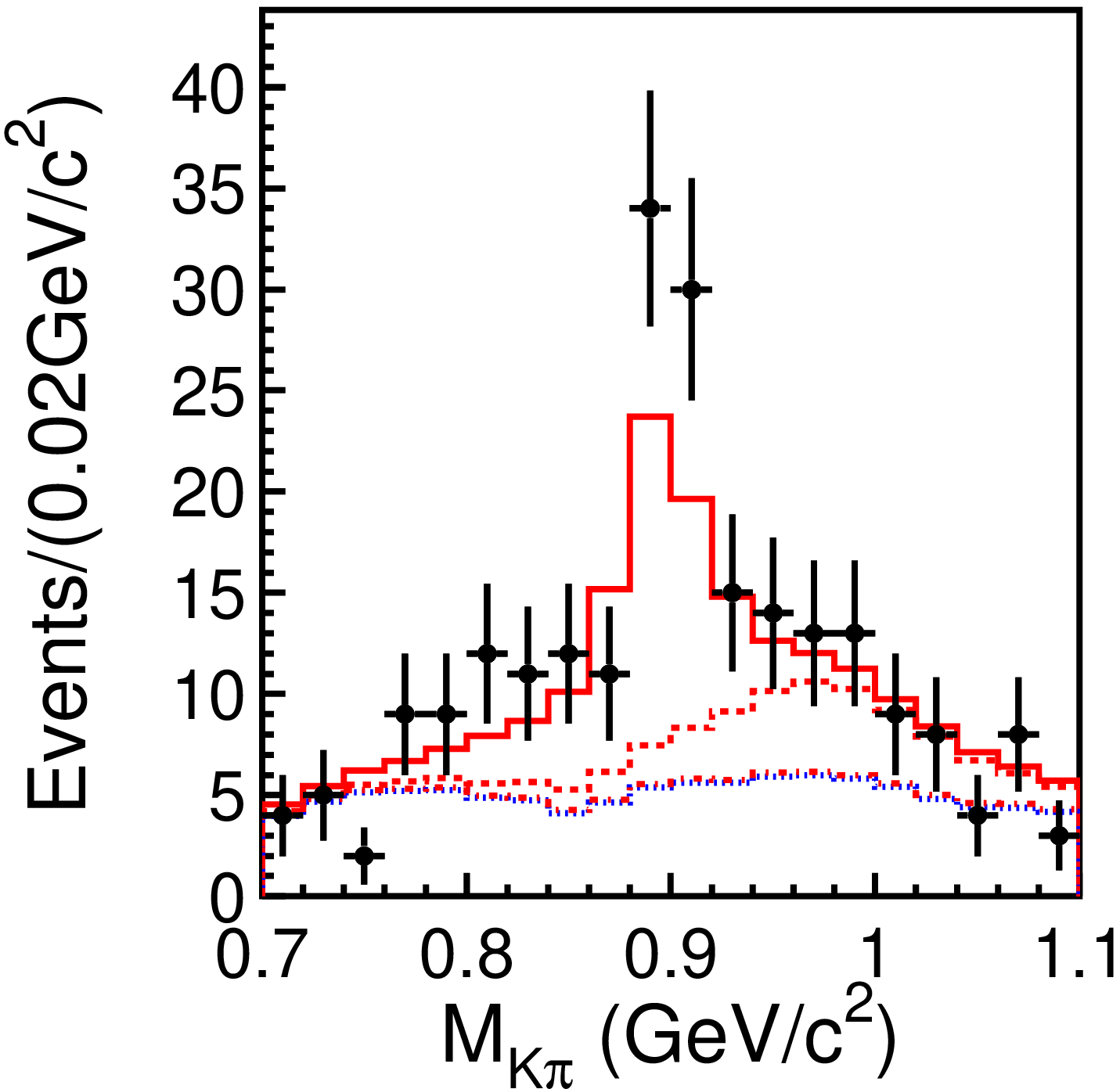}}
\end{tabular}
 \caption{$\dE$ (left) and $\mkpi$ (right) distributions for signal
 enhanced samples.
 The following selections are applied:
 $5.27\GeVcc\le\mb\le 5.29\GeVcc$ and $0.92\GeVcc\le\mkpi$ (left), and
 $5.27\GeVcc\le\mb\le 5.29\GeVcc$ and $-0.05\GeV\le\dE\le 0.1\GeV$ (right).
 Points with error bars are data. The solid
 histograms show the fit results. The areas divided by
 lines show the breakdown; from top to bottom, $\bz\to\kstarz\gamma$,
 $\bz\to\rhoz\gamma$, and other $\BB$ and $\qq$ components.
 Note that the other $\BB$
 component is too small to be visible in the plot on the right.}
 \label{fig:fsig}
\end{figure}

From the fit, we find $\Nrhozgam$ $\rhoz\gamma$ candidates,
$\Nkstarzgam$ $\kstarz\gamma$ background candidates,
$\NBB$ other $\BB$ background candidates, and
$\Nqq$ $\qq$ background candidates inside the signal box.
Figure~\ref{fig:fsig} shows the $\dE$ and $\mkpi$ projections of the fit
result for the signal enhanced samples.
The observed $\mkpi$ distribution is described well by our PDF,
which implies there is no significant contribution from non-resonant
$\pip\pim\gamma$ or $\kp\pim\gamma$.

We determine $\cals$ and $\cala$ from an UML fit to the observed $\Dt$
distribution.
For each event, the following likelihood function is
evaluated:
\begin{equation}
\begin{split}
 P_i =& (1-\fol)\int_{-\infty}^{+\infty} d(\Dt')
 \biggl[\sum_{j} f_j{\cal P}_j(\Dt')R_j(\Dt_i-\Dt')\biggr] \\
 +&\fol P_{\rm ol}(\Dt_i),
 \label{eq:likelihood}
\end{split}
\end{equation}
where $j$ runs over four components ($\bz\to\rhoz\gamma$,
$\bz\to\kstarz\gamma$, other $\BB$ and $\qq$).
The probability of each component ($f_j$) is calculated using the result
of the $\mb$-$\dE$-$\mkpi$ fit on an event-by-event basis. We also
incorporate the flavor tagging quality $r$ distribution. The $r$
distributions for $\kstarz\gamma$ and $\qq$ are obtained
by repeating the $\mb$-$\dE$-$\mkpi$ fit procedure to the signal sample
and also to the control sample for each $r$ interval
with yield parameters floated.
We found consistent distributions for the signal sample and the control sample.
The $r$ distribution for $\rhoz\gamma$ is expected to be consistent with
$\kstarz\gamma$, since the flavor is determined only by the tag side;
this is confirmed by MC.
The distribution of $\BB$ background is estimated from MC.

The PDF expected for the $\rhoz\gamma$ distribution,
${\cal P}_{\rhoz\gamma}$, is given by the time-dependent decay rate
[Eq.~(\ref{eq:psig})], modified to incorporate the effect of incorrect
flavor assignment; the parameters $\tau_\bz$ and $\dmd$ are fixed to
their world-average values~\cite{bib:PDG06}.  The distribution is then
convolved with the proper-time interval resolution function $\Rrho$,
which takes into account the finite vertex resolution.  The
parameterization of $\Rrho$ is the same as the one used in the
$\bz\to\phi\kz$~\cite{Chen:2006nk} analysis.
The same functional forms for the PDF and resolution are used for
the $\kstarz\gamma$ and other $\BB$ components, but
with separate lifetime and $CP$-violating parameters.  We assume no $CP$
asymmetry in $\kstarz\gamma$ and other $\BB$ background events; possible
deviations from this assumption are taken into account
in the systematic error. The
lifetime of $\bz\to\kstarz\gamma$ is the same as
$\bz\to\rhoz\gamma$. The effective lifetime of $\BB$ background is
obtained from a fit to the MC sample; the result is $\efflifer\,\rm ps$.
The PDF for $\qq$ background events, ${\cal P}_{\qq}$, is modeled as a
sum of exponential and delta function components, and is convolved with a double
Gaussian which represents the resolution function $\Rbkg$.  All
parameters in ${\cal P}_{\qq}$ and $\Rbkg$ are determined by a fit to
the $\Dt$ distribution in the $\dE$-$\mb$ sideband region ($\dE>0.2$
or $25(\mb-5.26)<(\dE-0.2)$ with $\dE$ in $\GeV$ and $\mb$ in $\GeVcc$).
$P_{\rm ol}$ is a Gaussian function
that represents a small outlier component with fraction $\fol$~\cite{bib:resol}.

The only free parameters in the $CP$ fit to $\bz\to\rhoz\gamma$ are
$\cals_{\rhoz\gamma}$ and $\cala_{\rhoz\gamma}$, which are determined by
maximizing the likelihood function 
 $L=\prod_iP_i(\Dt_i;\cals,\cala)$, where the product is over all events.
We obtain
\begin{eqnarray}
 \cals_{\rhoz\gamma} &=& \SrhozgmResultSS, \mbox{~and}\\
 \cala_{\rhoz\gamma} &=& \ArhozgmResultSS,
\end{eqnarray}
where the systematic errors are obtained as discussed below.

\begin{figure}
\begin{tabular}{cc}
 \resizebox{0.49\columnwidth}{!}{\includegraphics{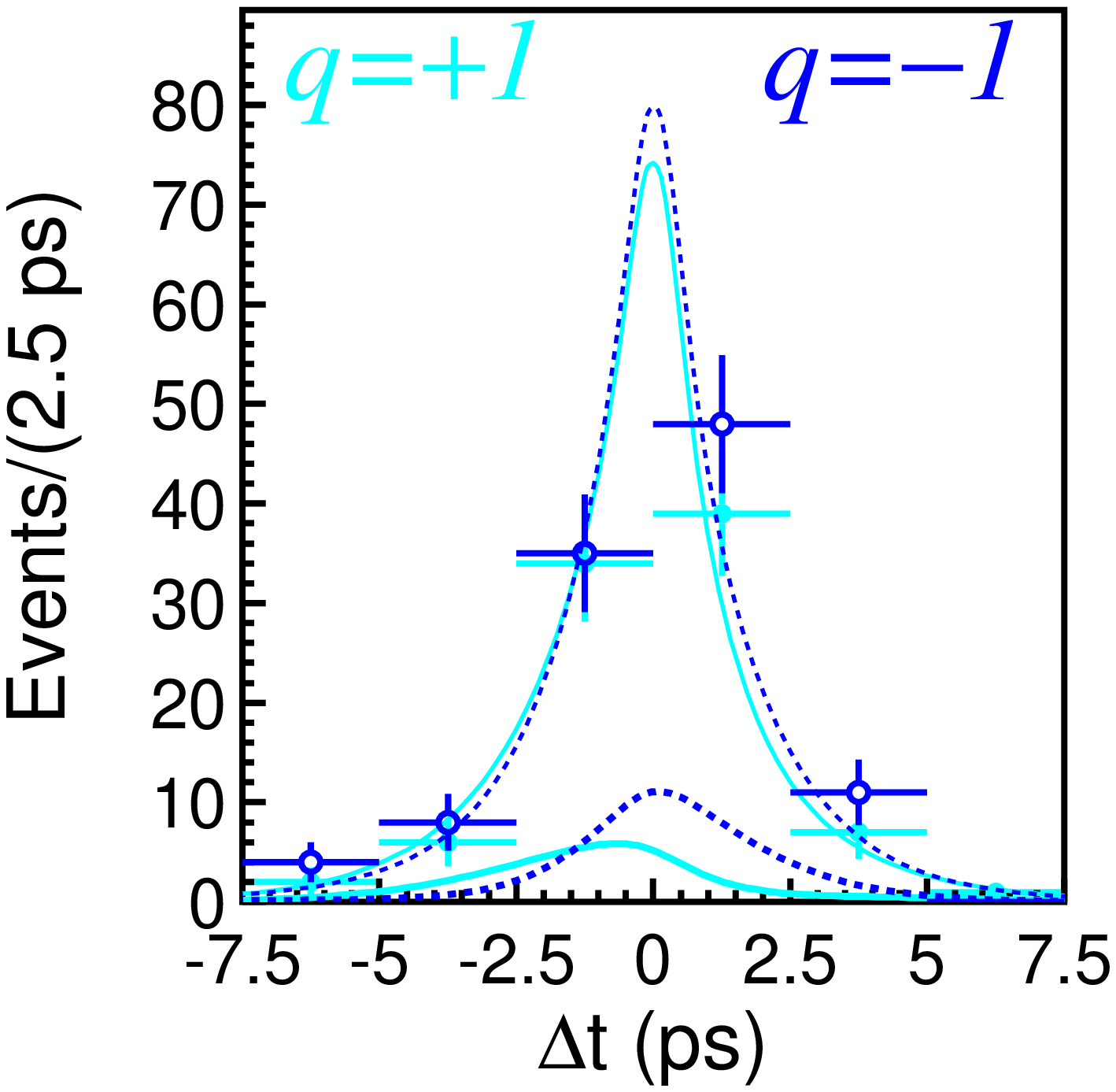}}
&
 \resizebox{0.49\columnwidth}{!}{\includegraphics{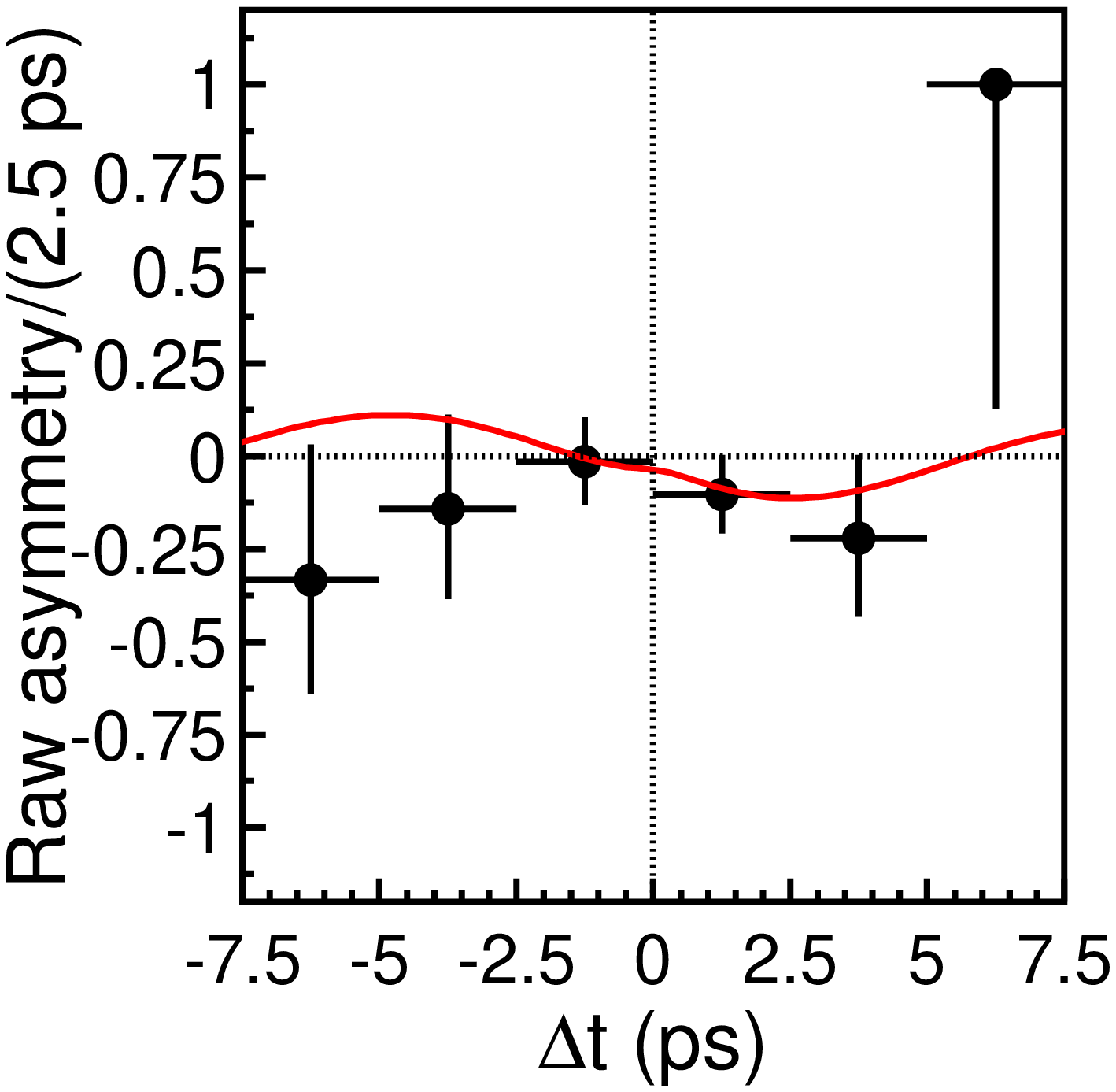}}
\end{tabular}
\caption{
 (Left) $\Dt$ distributions for $\bz\to\rhoz\gamma$
 for $q=+1$ (light solid) and $q=-1$ (dark dashed) with $0.5 < r \le 1.0$.
 The thin curve is the fit projection while the thick curve shows the
 signal component. Points with error bars are data.
 (Right) Raw asymmetry in each $\Dt$ bin with $0.5 < r \le 1.0$.
 The solid curve shows the result of the UML fit.
 }
\label{fig:asym}
\end{figure}

We define the raw asymmetry in each $\Dt$ bin by
$(N_{q=+1}-N_{q=-1})/(N_{q=+1}+N_{q=-1})$, where $N_{q=+1~(-1)}$ is the
number of observed candidates with $q=+1~(-1)$.  Figure~\ref{fig:asym}
shows the $\Dt$ distributions and the raw asymmetry for events with
$0.5 < r \le 1.0$.

We perform various validity checks of our fitting procedure.
A lifetime fit for the $\bz\to\kstarz\gamma$ control sample,
the $\kstarz\gamma$ component in the $\bz\to\rhoz\gamma$ sample
and the $\rhoz\gamma$ candidates gives $\controllife\,\rm ps$,
$\klife\,\rm ps$ and $\rlife\,\rm ps$, respectively. These results are all
consistent with the nominal $\bz$ lifetime ($\syme{1.530}{0.009}\,\rm ps$~\cite{bib:PDG06}).
A $CP$ asymmetry fit for the control sample gives an asymmetry
consistent with zero ($\cals = \SkstzgmResult$, $\cala =
\AkstzgmResult$).
A $CP$ asymmetry fit to the $\kstarz\gamma$ component in the
$\bz\to\rhoz\gamma$ sample also gives a consistent result
($\cals = \SkrResult$, $\cala = \AkrResult$).

We evaluate systematic uncertainties in the following categories by
fitting the data with each fixed parameter shifted by its $1\sigma$ error.
The largest contribution to the systematic error is from the uncertainty
in the probability of each component ($f_j$), because of the limited statistics;
we find an uncertainty of 0.16 on $\cals$ and 0.09 on $\cala$. 
The $CP$ asymmetry in $\kstarz\gamma$ has a direct impact on the
measurement. Based on the fit result from the control sample, we vary
$\cala_{\kstarz\gamma}$ from zero up to $\pm 0.05$, and find an error of
0.04 on $\cala$. The $CP$ asymmetry in other $\BB$ backgrounds has less
impact on the measurement.
This asymmetry is varied by the weighted average of possible maximum
$CP$ asymmetries ($\pm 1$ if not measured) of contributing decay modes
(0.06 on $\cals$, 0.09 on $\cala$);
we find an error of 0.01 or less on both $\cals$ and $\cala$.
The uncertainty from the resolution function parameters
is 0.06 on $\cals$ and 0.07 on $\cala$.
In addition to the above mentioned categories,
we also take the following small sources of uncertainty into account:
the uncertainty in the vertex reconstruction and flavor tagging,
uncertainty due to the tag-side interference effect~\cite{Long:2003wq}, 
uncertainty in the knowledge of the $\qq$ background $\Dt$ PDF,
uncertainty in the physics parameters such as $\dmd$, $\tau_\bz$,
possible effect of correlations between $\mb$, $\dE$ and $\mkpi$
and other possible biases.
Adding these contributions in quadrature,
we obtain a systematic error of 0.18 on $\cals$ and 0.14 on $\cala$.


In summary, we have measured the time-dependent $CP$ asymmetry in the
decay $\bz\to\rhoz\gamma$ using a sample of $\NBBoseven\,\BB$ pairs.  We
obtain $CP$-violation parameters
$\cals_{\rhoz\gamma} = \SrhozgmResultSS$ and
$\cala_{\rhoz\gamma} = \ArhozgmResultSS$.
With the present statistics,
the result is consistent with no $CP$ asymmetry and therefore no
indication of NP is found.  This is the first measurement of $CP$
asymmetry parameters in a $b\to d\gamma$ process.

We thank the KEKB group for excellent operation of the
accelerator, the KEK cryogenics group for efficient solenoid
operations, and the KEK computer group and
the NII for valuable computing and Super-SINET network
support.  We acknowledge support from MEXT and JSPS (Japan);
ARC and DEST (Australia); NSFC and KIP of CAS (China); 
DST (India); MOEHRD, KOSEF and KRF (Korea); 
KBN (Poland); MES and RFAAE (Russia); ARRS (Slovenia); SNSF (Switzerland); 
NSC and MOE (Taiwan); and DOE (USA).


\begin{thebibliography}{999}
\bibitem{Atwood:1997zr}
D.~Atwood, M.~Gronau and A.~Soni,
Phys.\ Rev.\ Lett.\  {\bf 79}, 185 (1997);

\bibitem{Ball:2006eu}
  P.~Ball, G.~W.~Jones and R.~Zwicky,
  Phys.\ Rev.\  D {\bf 75}, 054004 (2007).

\bibitem{Ushiroda:2006fi}
	Belle Collaboration, Y.~Ushiroda {\it et al.},
	Phys.\ Rev.\  D {\bf 74}, 111104 (2006).

\bibitem{Aubert:2005bu}
	\BaBar\ Collaboration, B.~Aubert {\it et al.},
	Phys.\ Rev.\ D {\bf 72}, 051103 (2005).

\bibitem{Mohapatra:2005rj}
	Belle Collaboration, D.~Mohapatra {\it et al.},
	Phys.\ Rev.\ Lett.\  {\bf 96}, 221601 (2006).

\bibitem{Aubert:2006pu}
	\BaBar\ Collaboration, B.~Aubert {\it et al.},
	Phys.\ Rev.\ Lett.\  {\bf 98}, 151802 (2007).

\bibitem{Ali:2004hn}
  A.~Ali, E.~Lunghi and A.~Y.~Parkhomenko,
  Phys.\ Lett.\  B {\bf 595}, 323 (2004);

  C.~D.~Lu, M.~Matsumori, A.~I.~Sanda and M.~Z.~Yang,
  Phys.\ Rev.\  D {\bf 72}, 094005 (2005)
  [Erratum-ibid.\  D {\bf 73}, 039902 (2006)].


\bibitem{Kim:2004zm}
  C.~S.~Kim, Y.~G.~Kim and K.~Y.~Lee,
  Phys.\ Rev.\  D {\bf 71}, 054014 (2005).

\bibitem{Belle}
  Belle Collaboration, A.~Abashian {\it et al.},
  Nucl. Instr. and Meth. A {\bf 479}, 117 (2002).
%
\bibitem{bib:KEKB}
  S.~Kurokawa and E.~Kikutani, 
  Nucl. Instr. and Meth. A {\bf 499}, 1 (2003)
	and other papers included in this Volume.
%


\bibitem{bib:CC}
        All decay modes include the charge conjugate, unless otherwise
        stated.

\bibitem{Koppenburg:2004fz}
Belle Collaboration, P.~Koppenburg {\it et al.},
Phys.\ Rev.\ Lett.\  {\bf 93}, 061803 (2004).

\bibitem{Fisher}
R.~A.~Fisher, Annals Eugen. {\bf 7}, 179 (1936).

\bibitem{Abe:2003yy}
Belle Collaboration, K.~Abe {\it et al.},
Phys.\ Rev.\ Lett.\  {\bf 91}, 261801 (2003).

\bibitem{bib:fbtg_nim}
H.~Kakuno, K.~Hara {\it et al.},
Nucl. Instr. and Meth. A {\bf 533}, 516 (2004).

\bibitem{bib:ARGUS}
ARGUS Collaboration, H.~Albrecht \textit{et al.}, 
Phys. Lett. B {\bf 241}, 278 (1990).

\bibitem{bib:PDG06}
Particle Data Group, W.-M.~Yao {\it et al.},
Journal of Physics G {\bf 33}, 1 (2006).

\bibitem{Chen:2006nk}
  Belle Collaboration, K.~F.~Chen {\it et al.},
  Phys.\ Rev.\ Lett.\  {\bf 98}, 031802 (2007).

\bibitem{bib:resol}
H.~Tajima {\it et al.},
Nucl. Instr. and Meth. A {\bf 533}, 370 (2004).

\bibitem{Long:2003wq}
O.~Long, M.~Baak, R.~N.~Cahn and D.~Kirkby,
Phys.\ Rev.\ D {\bf 68}, 034010 (2003).

\end{thebibliography}
\end{document}